\documentclass[10pt, conference, compsocconf]{IEEEtran}
\usepackage{url}
\usepackage{xcolor}
\usepackage{booktabs}
\usepackage{listings, multicol}

\ifCLASSINFOpdf
  \usepackage[pdftex]{graphicx}
\else
\fi

\usepackage{stfloats}
\begin{document}
%
\title{FirecREST: RESTful API on Cray XC systems}


\author{
\IEEEauthorblockN{
Felipe A. Cruz,
Maxime Martinasso
}
\IEEEauthorblockA{Swiss National Supercomputing Centre, ETH Zurich,
Lugano, Switzerland\\
 felipe.cruz@cscs.ch}
}


%


\maketitle

\begin{abstract}
As science gateways are becoming an increasingly popular digital interface for scientific communities, it is also becoming increasingly important for High-Performance Computing centers to provide a modern Web-enabled APIs. With such interface, science gateways can easily integrate access to HPC center resources. This work presents the FirecREST API, a RESTful Web API infrastructure that allows scientific communities to access the various integrated resources and services available from the Cray XC systems at the Swiss National Supercomputing Centre. FirecREST requirements are derived from use cases described in this work.

\end{abstract}


%
\IEEEpeerreviewmaketitle

\section{Introduction}
The recent technological progress on Web Application Programming Interfaces~\cite{masse2011rest} (Web APIs) provides third-party developers with frameworks for building HTTP based services that can be accessed by software applications over a variety of platforms. In this way, developers can envision and implement new business processes, build new client workflows that simplify the user experience or enable them to develop completely new platforms and services.

The core of the current Web API development gravitates towards Representational State Transfer (REST)~\cite{richardson2008restful}. RESTful API is a software design pattern, that specifies a uniform and predefined collection of stateless operations. The REST software architecture is well suited for enabling services that work over the Web as this design pattern introduces several desirable properties for web services, such as performance, scalability, and flexibility. As such, RESTful Web APIs provide interoperability between systems and applications over the web. It has become the building blocks for web software development, it reduces the total time required for software development of web-enabled applications and portals, and, it provides improved integration across multiple services and organizations.

Over the past year at the Swiss National Supercomputing Centre (CSCS), we developed FirecREST, a RESTful Web API infrastructure that allows scientific communities to access the integrated resources and services available at CSCS. FirecREST web services allow science gateway~\cite{doi:10.1002/cpe.3526} developers to integrate their platforms with High-Performance Computing (HPC) resources such as the CSCS flagship Cray supercomputer Piz Daint. In practice, the FirecREST API allows access to two core HPC services: submitting and monitoring jobs on HPC systems, and moving data across multiple filesystems. Moreover, it does so while enforcing integration with the Authorization and Authentication Infrastructure deployed at CSCS.

In this work, we present the architecture and the capabilities of the FirecREST API. Prior to describe the API, we introduce use cases that have driven its requirements. Whereas the work we present targets Cray systems, FirecREST is a generic interface that can be used on other non-Cray HPC systems. It interfaces primarily to the batch scheduler and HPC storage technology.

The key contributions of our work are:
\begin{itemize}
    \item to present concrete use cases that require the capability for an HPC center to provide resource access through a Web interface;
    \item to describe the architecture and capability of the FirecREST API.
\end{itemize}

\section{Use cases}

FirecREST will improve the accessibility of HPC resources to scientific communities by enabling them to build platforms that target specific scientific goals.
From the perspective of CSCS, providing a Web API to Cray system resources will expand the center user base as these scientific platforms develop. 
Platform developers will benefit from a standard interface, whereas CSCS provides a single technology to satisfy multiple user communities.
In this section, we present three use cases for which FirecREST provides key programmable functionalities.

\subsection{Swiss Data Science Center}

The Swiss Data Science Center\footnote{\url{https://datascience.ch}} (SDSC) aims to help the academic community and the industrial sector to work on Artificial Intelligence and Machine Learning, and, to facilitate the multidisciplinary exchange of data and knowledge.
SDSC has developed a solution called RENKU~\cite{renku}, a software platform for doing reproducible and collaborative data science.
 Piz Daint with its GPU-enabled nodes is a preferred infrastructure to run data science workload. Therefore, as requirements for FirecREST, users of RENKU should be able to submit jobs to Piz Daint and to move data to and from Piz Daint storage. As a software platform, RENKU requires an infrastructure to be executed on and both requirements need to be possible from any Cloud infrastructure.
RENKU is still in development and its integration with Piz Daint is in progress at the time of this writing.

\subsection{Materials Cloud}

The Materials Cloud\footnote{\url{https://www.materialscloud.org/}} is a web platform to share resources in computational materials science. One feature of the Materials Cloud is to run computational intensive jobs of well-known HPC-aware scientific application for discovering new properties of materials.

The Materials Cloud has been released and is online. It it is currently running on a Cloud system at CSCS. Computational intensive jobs are submitted to Piz Daint. To enable reproducible workflows and job submissions, the Materials Cloud uses AiiDA~\cite{PIZZI2016218}.
On the technical side, AiiDA interacts with Piz Daint by executing SSH commands.

AiiDA will greatly benefit from a RESTful API. It will ensure a better security, a lower effort of maintenance (as it is CSCS responsibility to provide a working API service) and a simplification of its internal mechanisms to access Piz~Daint. 

\subsection{Interactive CSCS service}

The third and last use case drives requirements for an internal CSCS service.
CSCS offers an interactive service\footnote{\url{https://jupyter.cscs.ch}} based on Jupyter notebooks~\cite{Kluyver:2016aa}.
Jupyter notebooks have become a preferred interface for scientists and many HPC centers provide a similar interactive service.
Moreover, science gateways are using Jupyter as a user-facing interface. For instance, the two previous use cases are running their web interface through a JupyterHub\footnote{\url{https://jupyter.org/hub}} service.

The current CSCS interactive service uses a standard Jupyter notebook spawner connected to the batch scheduler used to submit jobs on Piz Daint. In the future, this spawner will be modified to use a RESTful API. One of the main advantages of using a standard API technology is to easily target multiple infrastructures for executing notebooks. Each infrastructure has a cost model associated to, and, it becomes possible for the end users to target either an HPC system for dedicated resources and a higher cost or a Cloud system for shared resources and a lower cost.

\subsection{Summary and requirements}

From the above use cases we can identify three major requirements:
\begin{itemize}
    \item the necessity from the API to integrate with various identity providers external to the center. In the two first use cases, users don't necessarily have accounts at CSCS. This feature depends on the global Identity Management Access policy of the center and the use of standard authentication protocols;
    \item the capability to manage the execution of workloads on Piz Daint or any other HPC or Cloud systems;
    \item the possibility to enable external transfers of data to/from the centre filesystems attached to Piz Daint or any other HPC system.
\end{itemize}

We expect that FirecREST will enable new use cases and further increase the reach of HPC to scientific communities by: enabling the development of more modern and comfortable web interfaces to HPC, and, by providing an interface to the centre resources that is common, stable, secure, and maintainable, thus avoiding scientific community platforms to implement their custom interfaces and integration with infrastructure.

\begin{figure*}
  \centering
    \includegraphics[width=\textwidth]{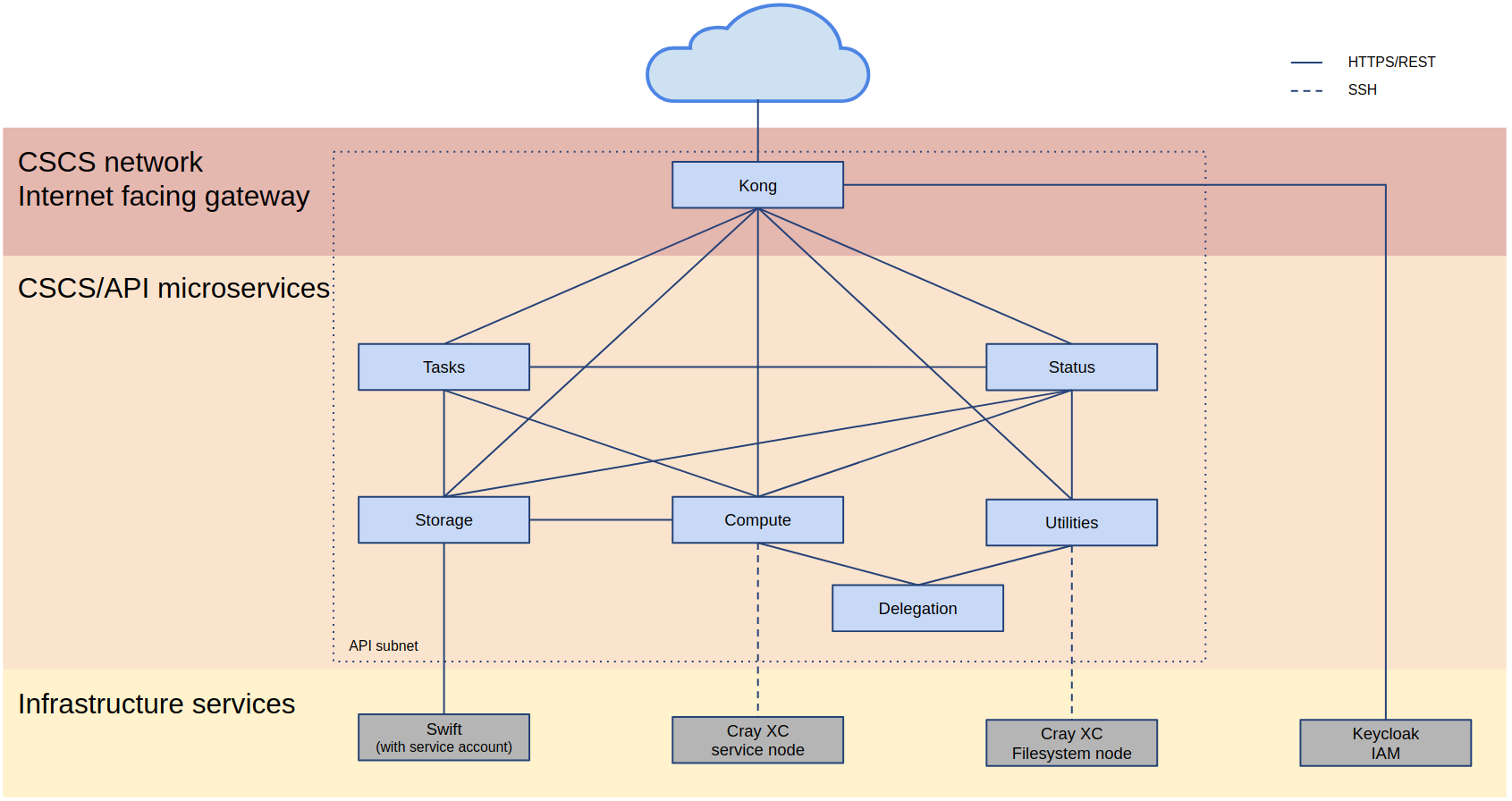}
  \caption{FirecREST microservice architecture.}
  \label{fig:workflowedge}
\end{figure*}

\section{FirecREST microservice architecture}

FirecREST provides to developers a web-enabled API to Piz Daint that is stable, secure, and maintainable. It allows client applications to access the resources available on the Cray XC system. Internally, FirecREST translates every HTTP request into its appropriate operations on the supercomputer, such as: enforcing authentication and authorization, job management, data mover, and other operations. The operations to perform are loosely coupled and involve different resources. Thus, in order to improve maintainability, test-ability, and to match CSCS organization, we followed a microservice architecture built using open source tools such as Keycloak\footnote{\url{https://www.keycloak.org/}}, Kong\footnote{\url{https://konghq.com/kong/}}, Flask\footnote{\url{http://flask.pocoo.org/}}, Paramiko\footnote{\url{http://www.paramiko.org/}}, OpenSSH\footnote{\url{https://www.openssh.com/}}, OpenAPI\footnote{\url{https://www.openapis.org/}}, and Redis\footnote{\url{https://redis.io/}}. Figure~\ref{fig:workflowedge} describes the FirecREST microservice architecture diagram.
In the following sub-sections, we present the core components and microservices that are part of FirecREST: Identity Access Management, API gateway, compute, storage, utilities, asynchronous tasks execution, delegation, and status.

\subsection{Identity Access Management}\label{sec:IAM}

The Identity and Access Management (IAM) infrastructure at CSCS ensures that users and web applications have the appropriate permissions to access resources at CSCS by using a secure protocol. From the whole of the IAM infrastructure at CSCS we will only discuss Keycloak, the Identity and Access Management solution deployed at the center. Keycloak allows to secure application and services by providing a mechanism for the authentication and authorization of CSCS users, CSCS services, and third party applications. Among the many features of Keycloak we highlight the following:

\begin{itemize}
\item single sign-on solution
\item integration with Kerberos authentication service
\item fine-grained authorization controls for services
\item client registration and authorization
\item support of OpenID Connect (OIDC) protocol\footnote{\url{https://openid.net/connect/}}
\end{itemize}

The integration of FirecREST with Keycloak has been achieved through the use of the OIDC protocol.
OIDC is an authentication protocol that extends the OAuth 2.0 specification. OAuth 2.0 is an industry-standard protocol for token-based authorization that is commonly used as a mechanism for users to grant access permission to web application in order to access user-owned resources and services.
The extensions provided by OIDC to the OAuth 2.0 protocol add a user authentication layer, providing a mechanism that enables single sign-on to users.

FirecREST leverages on Keycloak and OIDC for authentication and authorization of web applications, enabling the following capabilities:

\begin{itemize}
    \item enforce that all requests are authenticated
    \item applications never manipulate user credentials
    \item only allow requests from registered applications
    \item user-managed access permissions per application
    \item stateless security model by use of tokens
    \item short lifespan for sensitive access tokens
    \item extended client sessions allowed through refresh tokens
\end{itemize}

\begin{figure}
\centering
    \includegraphics[width=0.48\textwidth]{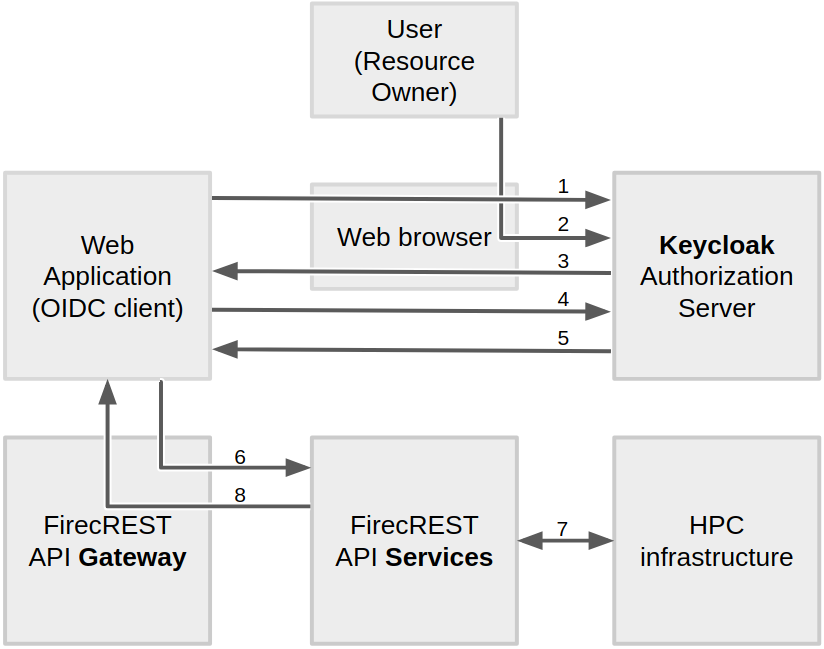}
  \caption{FirecREST authentication and authorization workflow: 1) Client performs a user authentication and access permission request; 2) User authenticates himself and authorizes client application; 3) Keycloak responds to the application with an authorization code; 4) Application uses the authorization code to request for an Access Token; 5) Keycloak grants the Access Token to the application over a secure backchannel; 6) Application request access to user-owned resources via the FirecREST API Gateway, enforcing the Access token permissions and redirecting to the correct service API endpoint; 7) FirecREST services translate the web request into actions on the HPC infrastructure; 8) FirecREST responds to the web application request.}
  \label{fig:firecrestAuth}
\end{figure}

In a nutshell, FirecREST OIDC-based IAM enables the user to login to a registered web application using their CSCS credentials and grant a web application with access to user-owned resource at the Center. Moreover, it does so without the user ever sharing their credentials with the web application. Figure~\ref{fig:firecrestAuth} presents a complete description of the OIDC Authorization Code Flow used by FirecREST.

\subsection{API Gateway}\label{sec:gateway}

The API gateway provides an interface to publish, maintain, monitor, and secure all the FirecREST API endpoints. As shown in Figure~\ref{fig:workflowedge}, the gateway is hosted on a machine within CSCS that is facing the internet. All interactions with the FirecREST API are first passed and validated before being redirected to any other FirecREST microservice.

In this way, every request made to the FirecREST API arrives first at the gateway, which will proxy the request towards the requested microservice endpoint. However, before the request is passed on to the microservice, the gateway will enforce that the request are correctly authenticated and authorized by requiring and validating the Access Token that must accompany each API request, as described in Section~\ref{sec:IAM}.

The current implementation of the gateway service is based on the Kong API gateway. Kong is a widely used open source microservice API gateway that implements functionalities such as a variety of authentication and authorization mechanisms, support for OIDC, IP filtering, access control lists, analytics, rate limiting, among many others that have allowed us to configure the gateway to our requirements.

\subsection{Compute}\label{sec:compute}

\begin{figure*}
\centering
    \includegraphics[width=0.98\textwidth]{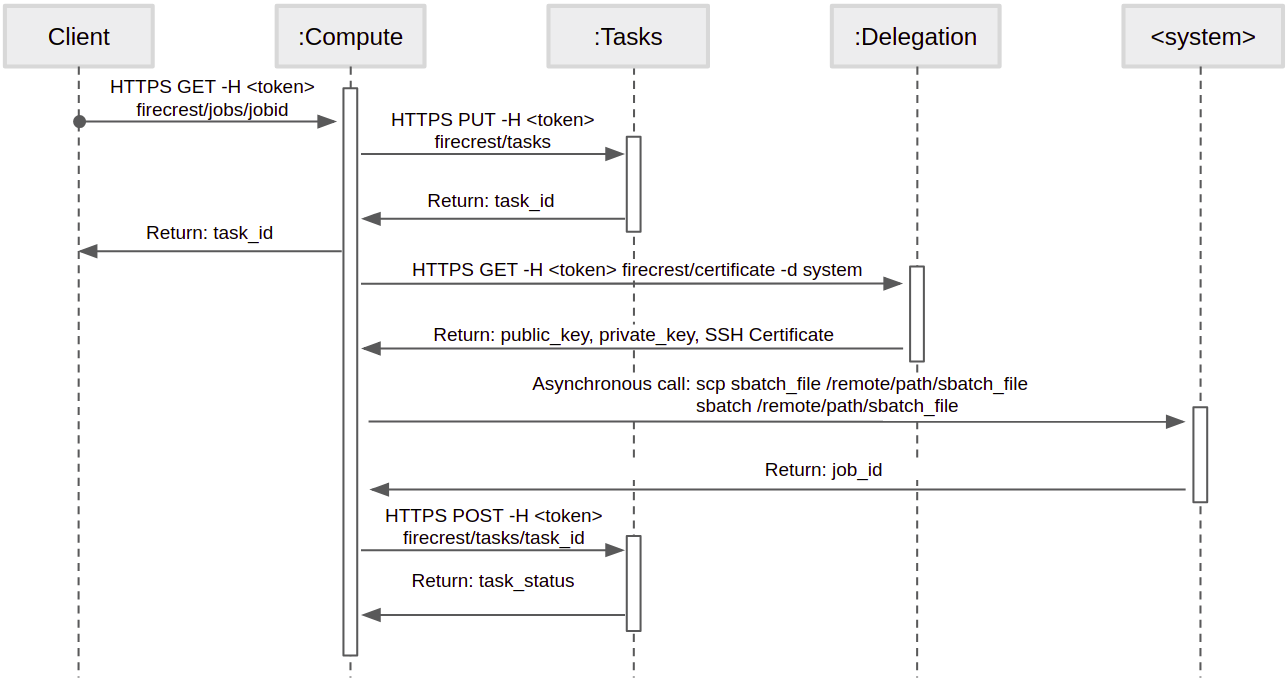}
  \caption{Sequence diagram of the job submission workflow. Please note that for conciseness we have skipped the API gateway step from the description, however, all interactions with the FirecREST API are passed and validated first by the gateway.}
  \label{fig:sequence_jobs}
\end{figure*}

The compute microservice implements the interface to the workload manager, thus allowing applications to submit, stop, and query the status of jobs by using non-blocking asynchronous API calls. This service depends on: the tasks microservice~(see section \ref{sec:tasks}) that provides a temporal resource that tracks the state of each call; and, the delegation microservice~(see section \ref{sec:delegation}) that issues a restricted SHH certificate that allows the execution of operations on behalf of the user. We now describe the integration with the SLURM~\cite{10.1007/10968987_3} workload manager used by Piz Daint.

For conciseness, in this section we will only describe the \textit{job submission} workflow, as other operations follow similar or simpler workflows.
Let us consider now the \textit{job submission} workflow shown in the sequence diagram in Figure~\ref{fig:sequence_jobs}.
As it can be observed, the job submission starts with the client calling the API endpoint \texttt{firecrest/jobs} with a \texttt{POST} operation, passing the following parameters: the access token which identifies the user and authorizes the call; the system where the job is submitted to; and, a file part that contains the job definition written in SLURM's sbatch format.
Upon receiving a request the compute microservice checks the validity of the parameters passed with the request and then call the tasks microservice, creating a new task which will track the progress of the operation returning a task resource as an immediate response to the client request. The client will use the task microservice to access the task resouce (identified by its \texttt{task id}) and use it to track the status of the request in an asynchronous way, meanwhile the compute microservice continuously updates the tasks resource as the job request progresses.

The compute microservice now requests to the delegation microservice an SSH certificate, passing the access token as a parameter. The delegation microservice will respond by retrieving the username from the access token and generating an SSH certificate that will be signed with the \textit{Certificate Authority key}~(see section \ref{sec:delegation} for details on this). Thus, the delegation microservice responds to the compute microservice request with a valid SSH certificate and the related public and private keys used in the process.

Next, the compute microservice makes use of the Paramiko library to establish an SSH session using the certificates and keys obtained from the delegation microservice. Over this SSH session, a unique temporal folder in the users's \texttt{\$HOME} directory is created, at this location the job information will be stored. The compute microservice then copies the sbatch script into the newly created temporal directory. Finally, the microservice runs the script using an sbatch command over the SSH session and captures its output, updating the task identified by the initial \texttt{task id} accordingly. Information in the task resource, such as \texttt{SLURM's job id field} can be used by the client to query for the state of the scheduled job.

We now provide an overview of all the compute microservice functionality by API endpoint, operation, and parameters:

\begin{enumerate}
    \item Endpoint: /Jobs/
    \begin{enumerate}
        \item Operation: POST
        \begin{itemize}
            \item Parameter: user, machine, scheduler script file.
            \item Description: Submits a job with the scheduler script file that targets the specified machine.
        \end{itemize}
        
        \item Operation: GET
        \begin{itemize}
            \item Parameter: user, machine.
            \item Description: Retrieve information from all jobs for the user at a specified machine.
        \end{itemize}
    \end{enumerate}
    
    \item Endpoint: /Jobs/acct
    \begin{enumerate}
        \item Operation: GET
        \begin{itemize}
            \item Parameter: user, machine.
            \item Description: Retrieves account information from user at specified machine.
        \end{itemize}
    \end{enumerate}
    
    \item Endpoint: /Jobs/jobid
    \begin{enumerate}
        \item Operation: GET
        \begin{itemize}
            \item Parameter: user, jobid, machine.
            \item Description: Retrieve information from a job with jobid at a specified machine.
        \end{itemize}
        \item Operation: DELETE
        \begin{itemize}
            \item Parameter: user, jobid, machine.
            \item Description: Cancels a job with jobid at a specified machine.
        \end{itemize}
    \end{enumerate}
\end{enumerate}

\subsection{Data mover}\label{sec:datamover}

\begin{figure*}
\centering
    \includegraphics[width=0.98\textwidth]{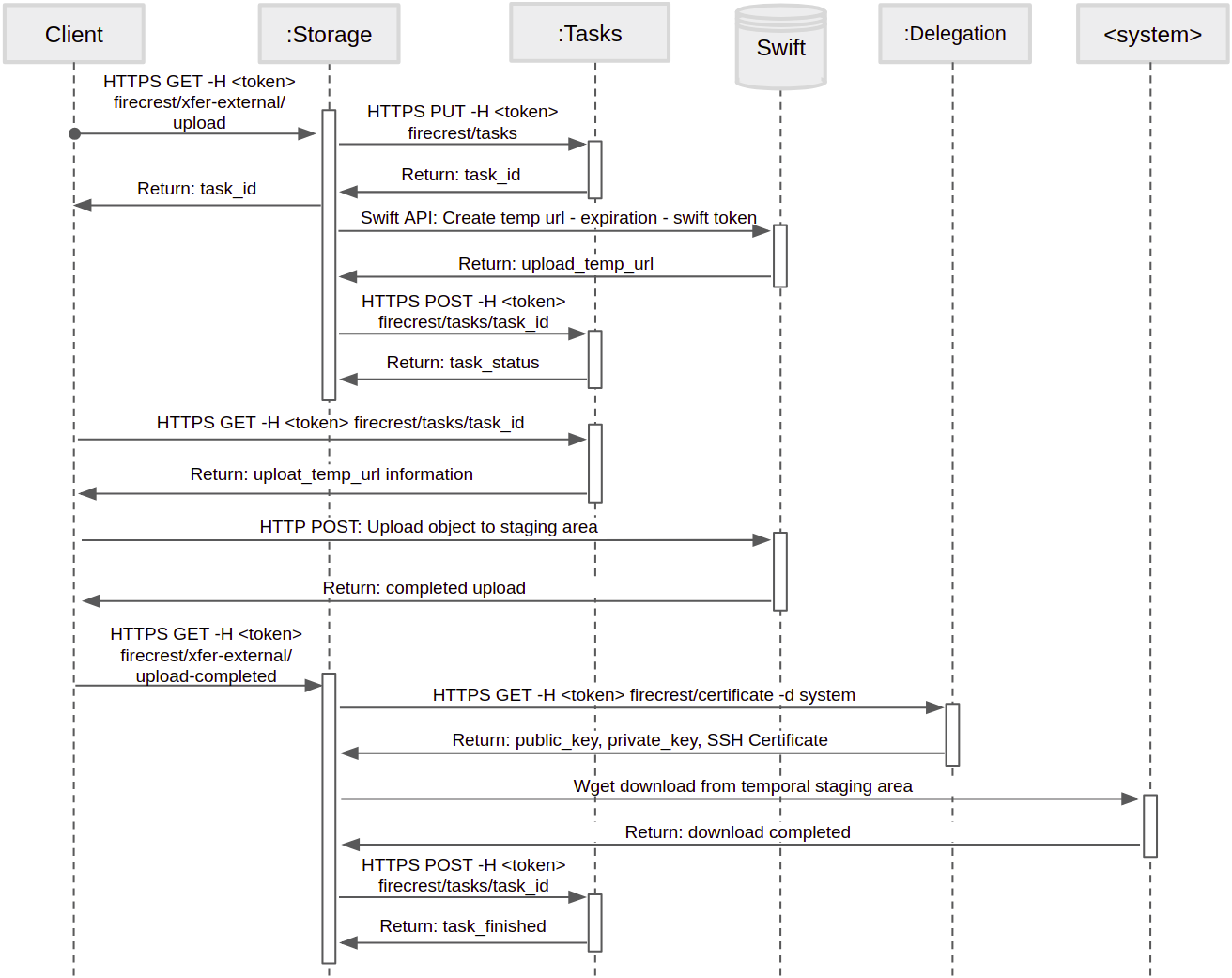}
  \caption{Sequence diagram of the data upload workflow that describes an asynchronous upload of a local file into CSCS infrastructure. The client first calls storage microservice giving parameters such as system, filesystem path and local file path. The storage microservice then uses Python's swiftclient with a FirecREST SWIFT service account in order to generate an HTTP form for temporary file upload. The client can then use the temporary upload URL to upload the file into an staging area. This circumvent the need for FirecREST to implement a high-performance fileserver, and leverages on existing solutions. Once the client upload to the staging area is completed, the user calls notifies the storage microservice by callong the upload-complete endpoint, this also provides the required user credentials that allow the storage microservice to start downloading object from SWIFT server into CSCS filesystem on behalf of the user.}
  \label{fig:sequence_upload}
\end{figure*}

\begin{figure*}
\centering
    \includegraphics[width=0.98\textwidth]{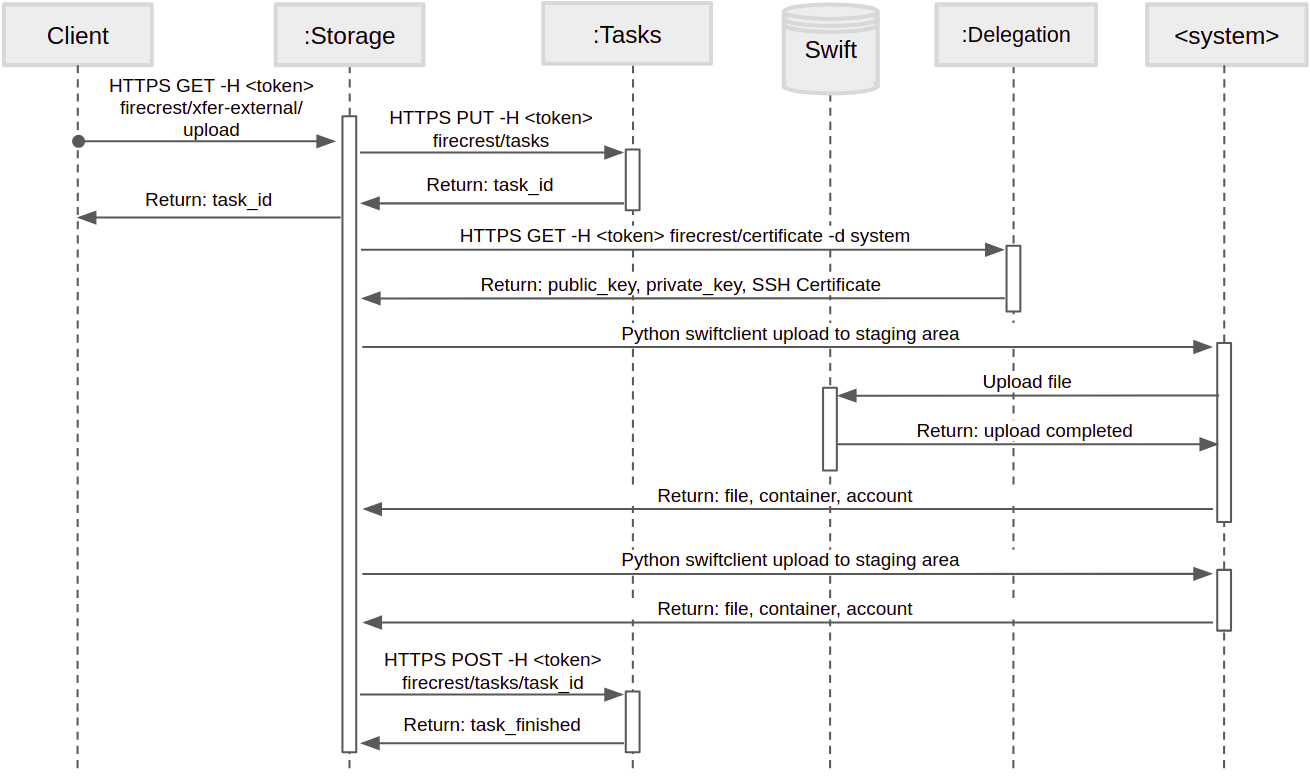}
  \caption{Sequence diagram of the data download workflow. The workflow for external data download is initiated by the client requesting a file from one of the HPC filesystem. An SSH user-certificate is created by the delegation microservice. The SSH certificate allows the storage microservice to upload the file into the FirecREST service account in SWIFT that is used as an staging area. Upon completion of the upload into the staging area, the storage microservice uses SWIFT API to create a temporary download URL. The temporal download URL is created by SWIFT with an unique hash containing that expiration time, object name and a secret, thus circumventing the need for the client to authenticate in order to start the remote download from SWIFT. Finally, temporary URL is returned to client.}
  \label{fig:sequence_download}
\end{figure*}

This microservice enable users the upload and download of large files to/from CSCS, while also enabling the movement of data within the different filesystems available on the Cray XC system (Piz Daint). It does so by using non-blocking calls to high-performance storage services while immediately responding with a reference to a resource that tracks the state of the request~(see section \ref{sec:tasks}). A full description of the upload and download workflow is presented in Figures~\ref{fig:sequence_upload} and Figure~\ref{fig:sequence_download} respectively.

We now provide an overview of all the storage microservice functionality by API endpoint, operation, and parameters:

\begin{enumerate}
    \item Endpoint: /Storage/xfer-external/upload
    \begin{enumerate}
        \item Operation: GET
        \begin{itemize}
            \item Parameter: user, target path.
            \item Description: First step of the asynchronous upload workflow, returns the tasks with information on the upload URL at the staging area.
        \end{itemize}
        \item Operation: GET
        \begin{itemize}
            \item Parameter: user, task-id.
            \item Description: Last step of the asynchronous upload workflow, moves data from upload staging area into POSIX filesystem.
        \end{itemize}
    \end{enumerate}
    \item Endpoint: /Storage/xfer-external/download
    \begin{enumerate}
        \item Operation: GET
        \begin{itemize}
            \item Parameter: user, source path.
            \item Description: Asynchronous download workflow, returns the tasks with information on the download URL from staging area once the file from the POSIX filesystem has been made available.
        \end{itemize}
    \end{enumerate}
    \item Endpoint: /Storage/xfer-external/{operation}
    \begin{enumerate}
        \item Operation: GET
        \begin{itemize}
            \item Parameter: user, operation, target path, source path (optional).
            \item Description: Asynchronous workflow for executing recursive operations on the POSIX filesystem, returns the tasks with job information scheduled to perform the operation. Operations that are possible: rsync, mv, rm.
        \end{itemize}
    \end{enumerate}
\end{enumerate}

\subsection{Utilities}\label{sec:utilities}

The utilities microservice provides synchronous execution of the following linux commands. As calls to the utilities microservice are blocking operations, these have a timeout and are not recursive.

\begin{itemize}
    \item GET /Utilities/ls
    \item GET /Utilities/file
    \item POST /Utilities/mkdir
    \item POST /Utilities/rename
    \item POST /Utilities/chmod
    \item POST /Utilities/chown
    \item POST /Utilities/symlink
\end{itemize}

The utilities microservice also provides two convenient endpoint for uploading and downloading smalls files, these transfers are limited to files under a few megabytes and are intended for setting up experiments and other limited filesystem updates.

\begin{enumerate}
    \item Endpoint: /Utilities/upload
    \begin{enumerate}
        \item Operation: POST
        \begin{itemize}
            \item Parameter: user, machine, path, file.
            \item Description: Blocking call that uploads a file to the specified path on the machine filesystem.
        \end{itemize}
    \end{enumerate}
    
    \item Endpoint: /Utilities/download
    \begin{enumerate}
        \item Operation: GET
        \begin{itemize}
            \item Parameter: user, machine, path.
            \item Description: Blocking call that returns the file from the specified path on the machine filesystem.
        \end{itemize}
    \end{enumerate}
\end{enumerate}

\subsection{Delegation}\label{sec:delegation}

The delegation microservice is a FirecREST internal service that is not exposed to the user. This service takes a valid JWT access token as input and creates a short-lived SSH certificate to be used to user authentication.

OpenSSH user-certificates are formed by: a public key; user identity information; and a set of constraints that limits the certificate validity. SSH certificates are signed using the \texttt{ssh-keygen} tool of \texttt{OpenSSH} with the \texttt{Certificate authority key} of the delegation microservice, also a standard SSH key. 
In order to enable SSH servers to accept certificates for user authentication, the sshd server must also be configured to trust the microservice CA public key.

Once a valid certificate is generated by the delegation microservice, other FirecREST microservices can use the certificate to perform remote command execution on behalf of the user. As such, this microservice enables FirecREST to perform delegation by means of secure system access using SSH certificates.

\subsection{Status}\label{sec:status}

This microservice provides information on the availability and state of services and relevant infrastructure that is accessible through FirecREST, such as the Cray XC system at CSCS.

We now provide an overview of all the status microservice functionality by API endpoint, operation, and parameters:

\begin{enumerate}
    \item Endpoint: /status/systems
    \begin{enumerate}
        \item Operation: GET
        \begin{itemize}
            \item Parameter: user.
            \item Description: Returns a list containing all available systems and response status.
        \end{itemize}
    \end{enumerate}
    
    \item Endpoint: /status/services
    \begin{enumerate}
        \item Operation: POST
        \begin{itemize}
            \item Parameter: user.
            \item Description: Returns a list containing all available microservices with a name, description, status, and endpoint.
        \end{itemize}
    \end{enumerate}
\end{enumerate}

\subsection{Tasks}\label{sec:tasks}

The task microservice responds to the need of managing the state of request that are being resolved asynchronously. One clear example for the need of this microservice can be observed in the data transfer operations handled by the storage microservice ~(see section \ref{sec:datamover}), as otherwise some of those workflows would not be possible. As such, firecrest microservices during an asynchronous request can rapidly create and respond with a new task resource. The operational result of the request is then tracked as the originating microservice continuously updates the task as progress is being made. Thus, task resources allow a client to perform other activities while a FirecREST asynchronous tasks are completed.

We now provide an overview of all the Tasks microservice functionality by API endpoint, operation, and parameters:

\begin{enumerate}
    \item Endpoint: /Tasks/
    \begin{enumerate}
        \item Operation: GET
        \begin{itemize}
            \item Parameter: user.
            \item Description: List all of the user's recorded tasks and their status.
        \end{itemize}
    \end{enumerate}
    
    \item Endpoint: /Tasks/
    \begin{enumerate}
        \item Operation: POST
        \begin{itemize}
            \item Parameter: user.
            \item Description: Create a new task entry to keep track and link to resources. Exclusively used by FirecREST microservices. Not exposed to user.
        \end{itemize}
    \end{enumerate}
        
    \item Endpoint: /Tasks/id
    \begin{enumerate}
        \item Operation: GET
        \begin{itemize}
            \item Parameter: user, id.
            \item Description: Retrieves information of a task. Exposed to user.
        \end{itemize}
        
    \item Operation: PUT
        \begin{itemize}
            \item Parameter: user, id.
            \item Description: Updates a tasks identified by its id. Exclusively used by FirecREST microservices. Not exposed to user.
        \end{itemize}
        
    \item Operation: DELETE
        \begin{itemize}
            \item Parameter: user, id.
            \item Description: Deletes a tasks identified by its id. Exclusively used by FirecREST microservices. Not exposed to user.
        \end{itemize}
    \end{enumerate}
\end{enumerate}

\section{FirecREST API specification}

The FirecREST API has been described using \texttt{OpenAPI}\footnote{\url{https://github.com/OAI/OpenAPI-Specification/}} in \texttt{YAML} format. \texttt{OpenAPI} is a language-agnostic standard for describing all aspects of a REST API: endpoints, endpoint operations, operation input parameters, operation output, and authentication methods. Moreover, the FirecREST project can leverage on open-source tools~\footnote{\url{https://swagger.io/docs/open-source-tools/swagger-editor/}}  built to support OpenAPI that simplify reading and writing the API, API documentation, and automatic library generation. 

%

\section{Related work}

The European project UNICORE~\cite{10.1007/3-540-44681-8_116}~\cite{romberg2000unicore} aims to develop a general-purpose federation software suite by following standard Grid and Web services. Its core framework also named UNICORE for Uniform Interface to Computing Resources can federate in a single view different systems ranging from high-end HPC systems to single Linux servers. UNICORE development follows a project-based funding which constraint the project to a discontinuous-pace development focusing on adding new features. For instance, key features such as using modern protocols for authentication and authorization or using a API specification like OpenAPI\footnote{\url{https://www.openapis.org/}} are not yet integrated. Our work offers a simpler approach by using a standard API which reduces its development cost compared to UNICORE.

NEWT~\cite{cholia2010newt}~\cite{cholia2015newt} aims to make HPC resources easily accessible to scientist by using Web applications. NEWT provides a RESTful API and we investigated its feature set prior to start this work. We found that some key aspects of our requirements were not integrated inside NEWT.
For instance, NEWT has a monolithic architecture with one point of failure, authentication and authorization should be ported to modern protocols to integrate with third party client and delegation needs to be implemented.
We concluded that NEWT has been developed and tailored with NERSC ecosystem requirements, and, porting it to CSCS environment requires an equivalent effort in terms of development. 


\section{Conclusion and future work}

In this work we present FirecREST, a RESTful Web API infrastructure that scientific gateways utilize to integrate with the High-Performance resources and services available from the Cray XC systems at CSCS. We intend to use FirecREST with the use cases presented in this paper.

As new use cases will emerge new requirements will be requested for FirecREST. As a concrete example, CSCS and the Paul Scherrer Institute (PSI) are collaborating to couple PSI scientific devices with CSCS compute capability. FirecREST is a key component to enable this connection, and, it will be extended to interface with a reservation service of compute nodes and a configurable data transformation service.







%
%
%

\bibliographystyle{IEEEtran}
\bibliography{CUG19_Firecrest}

\end{document}